\documentclass[12pt]{iopart}
\usepackage{epsfig,amssymb}
\def\beq{\begin{equation}}
\def\eeq{\end{equation}}
\def\bea{\begin{eqnarray}}
\def\eea{\end{eqnarray}}
\def\d{{\mathrm{d}}}

\newfont{\cursive}{pzcmi at 9pt}
\def\~t{\tilde{t}}

\begin{document}

\title{Conformally rescaled spacetimes and Hawking radiation}
\author{Alex B. Nielsen \\ Max-Planck-Institut f\"ur  
Gravitationsphysik, \\ Am M\"uhlenberg 1, \\ D-14476 Golm, Germany \\ {\it{and}} \\ Callinstrasse 38, \\ D-30167 Hannover, Germany \\ }
\author{J. T. Firouzjaee \\ Department of Physics, Sharif University of Technology \\ Tehran, Iran \\ {\it{and}} \\ School of Physics, \\ Institute for Research in Fundamental Sciences, \\ Tehran, Iran }

\begin{abstract} 
We study various derivations of Hawking radiation in conformally rescaled metrics. We focus on two important properties, the location of the horizon under a conformal transformation and its associated temperature. We find that the production of Hawking radiation cannot be associated in all cases to the trapping horizon because its location is not invariant under a conformal transformation. We also find evidence that the temperature of the Hawking radiation should transform simply under a conformal transformation, being invariant for asymptotic observers in the limit that the conformal transformation factor is unity at their location.
\end{abstract}


%\maketitle

%\tableofcontents

\section{Introduction}

Hawking radiation is an important prediction of black hole physics. In an otherwise empty space, quantum effects lead to a net radiation flux away from the black hole and this energy transfer implies that the black hole shrinks in size. It is one of the most studied predictions of the interplay of quantum fields with gravitation. Quantum effects lead to an evolution of the black hole that is radically different from that that would be derived from classical gravitational theory alone. There remain many open questions about Hawking radiation and critically an experimental confirmation. A similar effect is also expected for cosmological horizons \cite{Gibbons:1977mu}.

There have been many different derivations of the Hawking effect. The original Hawking calculation relied on calculating Bogoliubov coefficients for states that pass close to the black hole event horizon \cite{Hawking:1974sw}. Euclidean signature methods have been used to show that the Hawking temperature is related to the spacetime's periodicity in imaginary time \cite{Gibbons:1977mu}. Anomaly cancellation has been used by several authors to show the necessity of a Hawking-like flux at the quantum level; in 1+1 dimensions with the trace anomaly of the energy-momentum tensor \cite{Christensen:1977jc} and in four dimensional spherical symmetry for the gravitational anomaly in the divergence of the energy-momentum tensor \cite{Robinson:2005pd}. Another popular version is the tunneling calculation \cite{Parikh:1999mf,Vanzo:2011wq} that obtains the Hawking flux from a path integral across the horizon. The Hawking radiation is usually associated with the existence of an event horizon, but crucial in the tunneling calculation is a pole in the path integral that does not always occur at the location of the causal event horizon. This has led several authors to claim that Hawking radiation should be associated with quasi-local trapping horizons, both for black holes \cite{Di Criscienzo:2007fm} \cite{Hayward:2008jq} \cite{Nielsen:2008cr} and for cosmological horizons \cite{Cai:2008gw}.

The Hawking effect is a semi-classical quantum effect. The spacetime is treated classically but the matter fields are treated quantum mechanically. Previous results for static spacetimes have indicated that the surface gravity of a static black hole, derived purely classically, should be a conformal invariant, even if the conformal transformation is time dependent \cite{Jacobson:1993pf}. The equivalence of various other thermodynamic quantities under conformal transformations has been shown in \cite{Koga:1998un}. The conformal invariance of classical black hole thermodynamics has allowed conformal redefinitions to be used to calculate black hole thermodynamic properties in other generally covariant gravity theories \cite{Jacobson:1993vj}.

Most physical black holes will be dynamical to some extent. In dynamical spacetimes quasi-local horizons typically differ from null event horizons and other causal horizons. In recent work \cite{Nielsen:2010gm}, \cite{Faraoni:2011zy} the behaviour of quasi-local horizons under conformal transformations of the metric has been investigated. It was shown that the location of the trapping horizon is not conformally invariant. While it can appear only inside the event horizon in one conformal frame, it can appear partially outside in another \cite{Scheel:1994yn}. A number of authors \cite{Flanagan:2004bz}, \cite{Faraoni:2006fx}, \cite{Deruelle:2010ht} have argued, that if performed correctly, the outcomes of actual experiments should not be changed by a conformal transformation of the metric. Important here is that the conformal transformation should be accompanied by a change of measuring units \cite{Dicke:1961gz} that counterbalance the change of the spacetime metric.

The issue of the conformal equivalence of the Hawking effect poses an interesting challenge to claims that the trapping horizon is responsible for generating the Hawking flux. It was shown in \cite{Faraoni:2011zy} that the purely classical horizon-entropy increase law for quasi-local horizons could be preserved under a conformal transformation if the standard zero area expansion definition of the horizon is changed. Instead of a trapping horizon surface of zero instantaneous area expansion, one should consider a surface of zero entropy expansion. In this way the classical thermodynamic properties of quasi-local horizons can be maintained in agreement with the physical arguments of \cite{Flanagan:2004bz}, \cite{Faraoni:2006fx}, \cite{Deruelle:2010ht}.

We will study the issue of conformal transformations on Hawking radiation using several of the effective methods that have been used to study black hole radiation. These methods are necessarily heuristic since a fully back-reacting quantum evolution of black hole formation and evaporation from an initial quantum state is still missing. In particular the question of what vacuum state the initial quantum field should be in remains open \cite{Visser:2009pw}. In \cite{Marques:2011uq} the conformal invariance of the semi-classical Hawking effect was shown for a class of static, string-inspired spacetimes using a variety of different methods. In that work the location of the trapping horizon coincided with the event horizon in both conformal frames. We extend that work by considering conformal frames where crucially the locations do not coincide. 

Each of the semi-classical methods we study has its own strengths and weaknesses and range of applicability. The Euclidean section method is restricted to static spacetimes. Methods based on the calculation of Bogoliubov coefficients depend only on null, conformally invariant modes but some of these modes are trans-Planckian near the horizon. The tunneling methodology is most directly related to our purposes since it is predominantly this method that leads to the claim that it is the trapping horizon that produces the Hawking effect \cite{Di Criscienzo:2007fm}. The existence of a pole in the path integral at a certain spacetime point gives a direct localisation of the relevant horizon. The gravitational anomaly calculation is slightly more indirect, as it does not in itself locate a horizon and can be applied at different locations, but it is a curious fact that when applied to the quasi-local horizon it gives a thermal distribution, as opposed to its application on the event horizon or other related null surfaces. To our knowledge this is the first time that this observation has been made in the literature.

As concrete examples we will concentrate on the conformally rescaled Schwarzschild metric. This is largely for simplicity, and because the physical properties of the Schwarzschild spacetime are well studied. The first example is a static transformation of Schwarzschild studied in \cite{Deruelle:2010ht}. This is a simple static example with neat analytic solutions for the locations of various horizons, although the coordinates and conformal transformation are both singular at $r=2M$. We also discuss a static generalisation that is not singular at $r=2M$, but where a zero expansion horizon is still located outside of $r=2M$. The second example we look at is a time dependent conformal transformation of Schwarzschild that is regular on the $r=2M$ horizon. In this example, a future outer trapping horizon is located at $r=3M$.

We expect that our results can be generalised to other spacetimes but restrict ourselves in this work to spherical symmetry. Spherical symmetry has many practical advantages, a chief one being that the surfaces of spherical isometry define a natural slicing of the quasi-local horizons and so in turn select out ``preferred'' quasi-local horizons. This is despite the fact that even in spherical symmetry there exist many different intersecting quasi-local horizons \cite{Nielsen:2010wq}. We will ignore these and other technical issues in this work by focusing purely on spherically symmetric horizons.

The conformal transformations that we study are rather simple and by no means exhaust the full range of possibilities. Although conformal transformations do not change the local causal structure, conformal transformations that are not unity at infinity can change the global causal structure. For example, de Sitter spacetime is causally flat, but has a different global causal structure to Minkowski spacetime. A change of the full global causal structure is however not essential to our investigation. We take the position that observers do not need to be strictly located at asymptotic infinity to observe Hawking radiation - indeed in actual experimental realisations they will not be - and furthermore such observations by real observers will not necessarily encode information about inacessible asymptotic regions. One of the simple conformal transformations we consider is singular on the event horizon. Again this is a feature of its simplicity and not necessarily essential to our investigation. More complicated static conformal transformations that are regular on the event horizon could be investigated, and we give a simple example. Our interest is primarily in the behaviour near the quasi-local horizons and for that we need only the quasi-local conformally transformed geometry.

In the below we distinguish between $\omega_{K}$ being the ``invariant'' energy of \cite{hay0906} formed from the Kodama vector and $\omega$ being just the time derivative of the tunnelling action, $\partial_{t}I$, when the action is written $I= \int \omega\d t + \int k\d r$. Throughout we will use $l^{a}$ to denote the vector field of outgoing radial null vectors and $n^{a}$ to denote the corresponding ingoing vectors. Four-dimensional indices are denoted by Latin letters from the beginning of the alphabet, two dimensional indices by $i,j,k$ etc.

%********************************************************************************************************************************
\section{Conformal transformations of horizons}
%********************************************************************************************************************************

Consider a conformal transformation of an arbitrary spacetime metric, $g_{ab}$,
\beq \label{conformaltrans} g_{ab} \rightarrow \tilde{g}_{ab} = W(x)g_{ab} ,\eeq
where the conformal factor, $W(x)$, is a smooth, positive, non-zero, scalar function of the spacetime coordinates, which can include the time coordinate. We can assume that under the transformation the coordinates do not change, and an event in one conformal frame is labeled by the same coordinate values as it is in another conformal frame. Paths in the spacetime do not change. Timelike geodesics in one conformal frame are not necessarily geodesics in another conformal frame, but geodesic null rays remain geodesic, although not necessarily affinely parameterised.

The conformal factor changes the values of metric times and lengths. This can be compensated for by rescaling fiducial clocks and rods, the units of measurement \cite{Dicke:1961gz}. It has therefore been argued since the time of Dicke that such a transformation will not change the outcome of classical physical experiments performed in such a spacetime \cite{Flanagan:2004bz}. This is despite the fact that the transformation does change the geometry. Such a transformation can in certain cases be used to turn a curved space with non-vanishing Riemann tensor into a flat space with zero Riemann tensor (if the Weyl tensor vanishes). The transformation can also change a spacetime from one in which the energy conditions hold, into one in which the energy conditions are violated \cite{Magnano:1993bd}, but the physically measurable properties remain the same \cite{Faraoni:2006fx}.

Under a general conformal transformation the location of event horizons and all causal horizons, tied as they are to null rays and causality, remain unchanged. A Killing vector field, $k^{a}$, will remain a Killing vector field if the conformal factor, $W$, satisfies
\beq k^{a}\partial_{a}W = 0 \eeq
everywhere. In addition, a Killing horizon will remain a Killing horizon if
\beq \tilde{g}_{ab}k^{a}k^{b} = Wg_{ab}k^{a}k^{b} = 0 \eeq
at the Killing horizon location. If the first condition is not met then the conformal transformation will map a Killing horizon into a conformal Killing horizon \cite{Jacobson:1993pf}. The expansion, $\theta_{l}$, of an outgoing null vector field, $l^{a}$, changes as
\beq \theta_{l} \rightarrow \tilde{\theta}_{l} = \theta_{l} + \frac{l^{a}\partial_{a}W}{W} .\eeq
This change will affect the location of quasi-local horizons that are based on the vanishing of $\theta_{l}$, such as the apparent horizon or trapping horizon. For cases where $l^{a}\partial_{a}W \neq 0$ surfaces satisfying $\theta_{l}=0$ will not be at the same location as surfaces satisfying $\tilde{\theta}_{l}=0$. This variation cannot be compensated for by a rescaling of the null normals by $l^{a}\rightarrow \lambda l^{a}$, since such as transformation only changes the expansion by $\theta_{l} \rightarrow \lambda\theta_{l}$. The location of the $\theta_{l}=0$ surface is a coordinate invariant, although it is not a conformal frame invariant. A very simple example of this is the veiled Schwarzschild metric considered in \cite{Deruelle:2010ht};
\beq \label{veiledSchw} \d \tilde{s}^{2} = -\d t^{2} + \frac{\d r^{2}}{\triangle^{2}} + \frac{r^2}{\triangle}\d\Omega^{2} .\eeq
This can be obtained from the familiar Schwarzschild solution via the conformal transformation $g_{ab} \rightarrow \tilde{g}_{ab} = g_{ab}/\triangle$ where here $\triangle = 1-2M/r = 1/W$. It is argued in \cite{Deruelle:2010ht} that such a spacetime metric will have all the same physical predictions for physical features such as perihelion precession and Shapiro time delay as the familiar Schwarzschild solution under a corresponding variation of the physical units, as required by Dicke \cite{Dicke:1961gz}. In this transformed metric $M$ is just a parameter and should not necessarily be interpreted as a physical mass. The conformal factor is singular at the original horizon, $r=2M$, but it is a valid transformation in the region of outer communications, $r>2M$. 

For this veiled Schwarzschild case one finds $\theta_{l} =2(r-3M)/r^{2}$ up to a reparameterisation factor for $l^{a}$. The solution of $\theta_{l}=0$ is now at $r=3M$ and not at $r=2M$. This is despite the fact that there are causal curves that can cross $r=3M$ from $r<3M$ to $r>3M$. Because of the conformal factor, the $r$ coordinate is no longer the areal radius. The areas of the surfaces of spherical isometry are $4\pi r^{2}/\triangle$. This is an increasing function of $r$ for $r>3M$ but a decreasing function of $r$ for $r<3M$. There are thus spherical surfaces either side of $3M$ that have the same area. It is tempting to recast this solution in terms of the areal radius $R=r/\sqrt{\triangle}$ but this is not a good coordinate at $r=3M$.

This surface at $r=3M$ is somewhat unusual in that $\theta_{n}$ is also zero there and it is a timelike surface, not a null surface. It is neither an isolated horizon in the sense of \cite{Ashtekar:2003jh} nor a future outer trapping horizon in the classification of \cite{Hayward:1993wb}, but this is a feature of the simple example rather than generically expected behaviour.

In this example, the singular $r=2M$ surface is also not a Killing horizon, although the metric remains static everywhere and the Killing vector field unchanged. A simple example that retains the Killing horizon at $r=2M$ but shifts the $\theta_{l}=0$ horizon to $r=5M/2$ is obtained by choosing $W=1/\sqrt{\triangle}$. An example that is regular at the horizon and everywhere in the domain of outer communications but still asymptotically flat can be obtained by choosing a conformal factor of the form $W=r/(r-aM)$ with $a$ slightly less than $2$. In this case the outgoing expansion is $\theta_{l}=\triangle(2r-3aM)/(r-aM)r$ and vanishes at both $r=2M$ and $r=3aM/2$.

A fully regular and static conformal transformation of the Schwarzschild solution, as is for example considered in \cite{Marques:2011uq}, will not change the location of the Killing horizon, nor will it change the location of the $\theta_{l}=0$ horizon at $\triangle =0$. In general though, for suitable choices of parameters, it is possible for a regular, dynamical conformal transformation to change the location of $\theta_{l}=0$ surfaces, such that none of the new $\theta_{l}=0$ surfaces will be at $\triangle=0$ and some of them will be future outer trapping horizon in the classification of \cite{Hayward:1993wb}. To illustrate this and emphasise the regularity at $\triangle=0$ we can adopt horizon regular advanced Eddington-Finkelstein coordinates. Under a conformal transformation, a general spherically symmetric metric in advanced Eddington-Finkelstein coordinates will be
\beq \label{conftransEF} \d \tilde{s}^{2} = -W(v,r)A^{2}(v,r)\triangle(v,r)\d v^{2} + 2W(v,r)A(v,r)\d v\d r + W(v,r)r^{2}\d\Omega^{2} ,\eeq
where $\triangle(v,r)$ is a general functions of $v$ and $r$ and we require $W(v,r)>0$ and $A(v,r)>0$ everywhere. This is just a conformal rescaling of the familiar spherically symmetric Eddington-Finkelstein metric if $A=1$ and $\triangle = 1-2M/r$. In these horizon regular coordinates $(v,r,\theta,\phi )$, future directed radial null vectors are given by
\beq l^{a} = \left( 1, \frac{A\triangle}{2}, 0 ,0 \right) ,\eeq
and
\beq n_{a} = \left( -1, 0, 0 ,0 \right) .\eeq
For the restricted case where the conformal factor is only time dependent, $W(v)$, the radial null expansions take a rather simple form. In this case they are
\beq \theta_{l} = \frac{\triangle A}{r} + \frac{\partial_{v} W}{W} ,\eeq
and
\beq \theta_{n} = -\frac{2A}{rW} .\eeq
In addition one finds
\beq n^{a}\nabla_{a}\theta_{l} = \frac{\triangle(A-r\partial_{r}A) - Ar\partial_{r}\triangle}{WAr^{2}} .\eeq
For $\partial_{v}{W}\neq 0$ the marginally outer trapped surfaces do not lie at $\triangle = 0$, as they would do if $W$ were everywhere unity. 

Since we require $W>0$, we will always have $\theta_{n}<0$. We then need $\partial_{v} W<0$ to get a solution of $\theta_{l}=0$ at positive $\triangle$ and $r$. With $A=1$ and $\triangle = 1-2M/r$, $M$ constant, we have $\theta_{l}=0$ at
\beq r = \frac{1}{2}\frac{W}{\partial_{v} W}\left(-1 \pm \sqrt{1+8M\partial_{v} W/W}\right) .\eeq
As a concrete example of this choose
\beq W=\mathrm{exp}(-v/9M).\eeq
such that the metric takes the form,
\beq \label{timedepmetric} \d s^{2} = -e^{-v/9M}\triangle\d v^{2} + 2e^{-v/9M}\d v\d r + e^{-v/9M}r^{2}\d\Omega^{2} .\eeq
Then we will have horizons at $r=3M$ and $r=6M$ with an untrapped region between them since the expansions are given by
\beq \theta_{l} = -\frac{1}{9Mr^{2}}(r-3M)(r-6M) ,\eeq
\beq \theta_{n} = -\frac{2e^{v/9M}}{r} ,\eeq
\beq n^{a}\nabla_{a}\theta_{l} = \frac{(r-4M)e^{v/9M}}{r^{3}} .\eeq
The surface at $r=3M$ is a future outer trapped horizon (FOTH) in the classification of \cite{Hayward:1993wb}. The resulting metric is asymptotically flat on constant $v$ slices, but this is not critical for our purposes and not relevant to finding the trapping horizons. One could imagine a slightly more complicated conformal factor $W(v,r)$ that drops off to one outside some region of interest, far from $r=2M$, such that the resulting spacetime is asymptotically Minkowski. Other examples, including cosmological examples, are discussed in \cite{Faraoni:2011zy}.

In \cite{Nielsen:2010gm} it was shown that a conformally invariant quasi-local horizon can be defined in terms of the vanishing change of the Wald entropy \cite{Wald:1993nt} in the outgoing null direction. If the examples given above are viewed as solutions of the conformally transformed Einstein equations then these quasi-local horizons will be located at $r=2M$, not $r=3M$. Furthermore, it was shown that at the purely classical level, an entropy increase law can be derived for these conformally invariant quasi-local horizons that is analogous to the second law of thermodynamics. In the present work we wish to investigate to what extent the quantum effect related to black hole thermodynamics, the Hawking radiation, can also be associated with conformally invariant horizons.

%****************************************************************************************************************
\subsection{Misner-Sharp mass }
%****************************************************************************************************************

An important condition for the location of the spherically symmetric trapping horizon in spherical symmetry is the condition that the Misner-Sharp mass, $M_{_{\mathrm{MS}}}$, should be equal to one half of the areal radius coordinate, $R$, chosen such that the area of surfaces of spherical isometry is $4\pi R^{2}$. The area of surfaces of spherical isometry changes under a conformal transformation implying that the areal radius changes too. But the Misner-Sharp mass is also changed, and the condition $R=2M_{_{\mathrm{MS}}}$ remains satisfied at the shifted trapping horizon location.

In \cite{Misner:1964je} the Misner-Sharp mass is defined for the following metric
\beq\label{general-m} ds^{2}=-e^{2\nu(t,r)}dt^{2}+e^{2\psi(t,r)}dr^{2}+R(t,r)^{2}d\Omega^{2}, \eeq
by
\beq \label{mtt} M_{_{\mathrm{MS}}}= \frac{R}{2}\Bigg(1-(G-H)\Bigg),\eeq
where
\beq G\equiv e^{-2\psi}(\partial_{r}R)^{2}~~,~~H\equiv e^{-2\nu}(\partial_{t}R)^{2}.\eeq
The Misner-Sharp mass will equal half the areal radius if $G-H=0$. This combination is simply
\beq G-H = \gamma^{ij}\nabla_{i}R\nabla_{j}R \eeq
where $\gamma^{ij}$ is the inverse of the two-dimensional metric of the $t$ and $r$ coordinates. In \cite{DiCriscienzo:2010zza} this combination is given the symbol $\chi$ and it is related to the radial null expansions by
\beq \chi \equiv \gamma^{ij}\nabla_{i}R\nabla_{j}R = -\frac{R^{2}}{2}\theta_{l}\theta_{n} .\eeq
The areal radius, $R$, is not conformally invariant because the area of spheres is not conformally invariant. Similarly, the Misner-Sharp mass will not take the same value at the same point in a conformally transformed metric as it does in the original metric. It will equal half the areal radius when either of the expansions vanish, but this will not occur at the same coordinate values as in the original metric.
For the static conformally transformed Schwarzschild example (\ref{veiledSchw}) we have
\beq e^{2\nu(t,r)}=1,~~e^{2\psi(t,r)}=
\frac{1}{\Delta^2},~~R^2=\frac{r^2}{\Delta} . \eeq
The parameter, $M$ is not the Misner-Sharp mass in (\ref{veiledSchw}). The Misner-Sharp mass for this metric is,
\beq M_{_{\mathrm{MS}}}=\frac{M}{2\sqrt{1-\frac{2M}{r}}}
\left(\frac{4r-9M}{r-2M}\right) ,\eeq
which is singular at $r=2M$ and negative for $2M<r<9M/4$. The $r=3M$ surface is however the $R=2M_{_{\mathrm{MS}}}$ surface for this metric, as expected.

%********************************************************************************************************************************
\section{The classical surface gravity}
%********************************************************************************************************************************

It was shown in \cite{Jacobson:1993pf} that under a conformal transformation that is unity at infinity, the classical surface gravity of a Killing horizon is a conformal invariant. The method used is purely classical with a surface gravity, $\kappa$, defined by
\beq \label{confsurfgrav} \nabla_{a}\left(g_{bc}\chi^{b}\chi^{c}\right) = -2\kappa g_{ad}\chi^{d} .\eeq
$\chi^{a}$ is a vector field that at spacelike infinity satisfies $\chi^{a}\chi_{a}=-1$ and the equation is evaluated where $\chi^{a}\chi_{a}=0$. Since the conformal factor is unity at infinity, this normalisation is preserved and the conformal transformation just maps $\chi^{a}$ to $\chi^{a}$. Under a general conformal transformation of the form (\ref{conformaltrans}) one then obtains
\bea  \tilde{\nabla}_{a}\left(\tilde{g}_{bc}\chi^{b}\chi^{c}\right) & = & \nabla_{a}\left(Wg_{bc}\chi^{b}\chi^{c}\right) \nonumber \\
& = & \left(\nabla_{a}W\right)g_{bc}\chi^{b}\chi^{c} + W \nabla_{a}\left(g_{bc}\chi^{b}\chi^{c}\right) \nonumber \\
& = & -2\kappa W g_{ad}\chi^{d} \nonumber \\
& = &  -2\kappa \tilde{g}_{ad}\chi^{d} ,\eea
hence the surface gravity is conformally invariant, $\kappa = \tilde{\kappa}$, with this definition. This is true even if the conformal transformation turns a Killing horizon into a conformal Killing horizon, for example in the case where a non-static conformal transformation is imposed on a static spacetime. For restricted conformal transformations that map true Killing horizons into true Killing horizons (as above when $k^{a}\partial_{a}W = 0$) then other definitions of the surface gravity, such as $\chi^{a}\nabla_{a}\chi^{b} = \kappa\chi^{b}$ are also conformally invariant.

In light of the claimed physical equivalence of conformal frames, the fact that some definitions of surface gravity display the full conformal invariance, whereas others only display it for certain restricted transformations, means it may be possible to use this to prefer certain definitions, especially when extending to other situations of non-Killing horizons. Furthermore, since the Hawking temperature is expected to be related to the surface gravity by $T=\kappa/2\pi$ the conformal invariance of the surface gravity suggests that the Hawking temperature should also be conformally invariant. Since this temperature refers to the temperature measured by inertial observers at infinity, the requirement that the conformal factor be one at infinity ensures that the units of temperature do not need to be scaled at infinity and the outcome of a temperature measuring experiment remains invariant. If one normalises the Killing vector for a stationary observer location, $r_{obs}$, not at infinity, such that $\chi^{a}\chi_{a}=-1$ at $r_{obs}$, then the surface gravity scales as $\tilde{\kappa}_{obs} = \kappa_{obs}/\sqrt{W_{obs}}$, which scales exactly like a mass or an energy and is compensated for by a shift in fiducial units \cite{Dicke:1961gz}.

The Kodama surface gravity, $\kappa_{K}$, can be computed using \cite{Hayward:1997jp}
\beq \kappa_{H} = \frac{1}{2\sqrt{-\gamma}}\partial_{i}\left(\sqrt{-\gamma}\gamma^{ij}\partial_{j}R\right) .\eeq
For the static veiled Schwarzschild example (\ref{veiledSchw}) this computes to
\beq \kappa_{H} = \frac{r}{4}\left(\frac{(\triangle ')^{2}}{2\sqrt{\triangle}}-\sqrt{\triangle}\triangle ''\right) .\eeq
Its value at $r=3M$ is $1/(6\sqrt{3}M)$ and it diverges in the limit $\triangle \rightarrow 0$ or $r \rightarrow 2M$. Under a conformal transformation (\ref{conformaltrans}) the Kodama surface gravity transforms as
\beq \kappa_{K} \rightarrow \tilde{\kappa}_{K} = \frac{\kappa_{K}}{\sqrt{W}} + \frac{1}{2W\sqrt{-\gamma}}\partial_{i}\left(\sqrt{-\gamma}\gamma^{ij}R\partial_{j}\sqrt{W}\right) ,\eeq
when the areal radius transforms as $R \rightarrow \tilde{R} = \sqrt{W}R$ and the two-dimensional normal metric $\gamma_{ij} \rightarrow {\tilde{\gamma}}_{ij} = W\gamma_{ij}$. Thus it does not enjoy the same conformal invariance as (\ref{confsurfgrav}), even for the case where $W$ is asymptotically unity and {\it{regular}} on the horizon. For the example with $W=r/(r-aM)$ and $a<2$, even at $\triangle=0$ its value is
\beq \tilde{\kappa}_{K} = \frac{\kappa_{K}}{\sqrt{W_{H}}} - \frac{1}{\sqrt{W}}\frac{1}{4r_{H}}\frac{a}{2-a} \eeq
which is not a simple conformal transformation of an energy or temperature. This is perhaps not fatal in itself as the physically relevant quantity in \cite{hay0906} is a ratio of the Kodama surface gravity and the energy and the Kodama surface gravity is not itself a physical observable. But it demonstrates that the Kodama surface gravity behaves differently to textbook definitions such as (\ref{confsurfgrav}) and that it will not be straightforward to adapt the method of \cite{Jacobson:1993vj} to use the Kodama formalism in non-Einstein theories of covariant gravity.

The behaviour of thermodynamic properties of static horizons under conformal transformations was also examined in \cite{Koga:1998un}. It was found that other thermodynamic properties are also conformally invariant when defined on a static Killing horizon. These results however are only valid for static horizons and only use purely classical methods for calculating the relevant thermodynamic quantities. The semi-classical case was investigated in \cite{Marques:2011uq}, for a class of string-inspired models and transformations that preserve the Killing horizon property of the horizon. They investigated a number of different techniques for deriving the Hawking effect semi-classically; the gravitational anomaly, the metric Euclideanization, the Bogoliubov coefficients and the reflection coefficient. They found for the collection of methods used, and a conformal transformation that maps Killing horizons into Killing horizons, that the black hole temperature remains invariant.

We turn now to applying these techniques more generally to see what effect general conformal transformations have on the Hawking effect for general spherically symmetric spacetimes.

\section{Euclidean section method}
%****************************************************************************************************

In static spacetimes a temperature can be derived by looking for periodicity in the imaginary Euclidean time \cite{Gibbons:1977mu}. With the transformation $\tau = it$, the Euclidean signature version of a static, spherically symmetric metric with $A(r)$ and $\triangle(r)$ is
\beq \d s^{2} = A^2 \triangle \d \tau^{2} + \frac{\d r^{2}}{\triangle} + r^{2}\d\Omega^{2} .\eeq
Under this Wick rotation, for fixed angular coordinate values, the horizon is mapped to the origin of a two dimensional flat surface. In these coordinates there is still an apparent coordinate singularity at $\triangle = 0$. We introduce the new coordinate $x=\alpha\sqrt{\triangle}$ with $\alpha$ a constant. (This is not the tortoise coordinate $\d r_{*} =\d r/\triangle$.) The two-dimensional metric becomes
\beq \d s^{2} = x^{2}\left(\frac{A}{\alpha}\right)^2 \d\tau^{2} + \frac{4}{\alpha^{2}\partial_{r}\triangle^{2}}\d x^{2} .\eeq
We can now arrange for the $\tau, x$ coordinates to correspond to polar coordinates $R$ and $\Theta$ of a flat metric $R^{2}\d\Theta^{2} + \d R^{2}$ at the horizon. To do this we must pick
\beq \alpha = \frac{2}{(\partial_{r}\triangle)_{H} } \eeq
where the subscript $H$ denotes evaluation of the function at $\triangle=0$. This choice makes the $x$ coordinate become the $R$ coordinate and the $\tau$ coordinate the $\Theta$ coordinate. In this case $A\tau/\alpha$ must have period $2\pi$ in order for there to be no conical singularity at the horizon. This implies that $\tau$ should have period $\beta = 2\pi\left(\frac{A_{H}(\partial_{r}\triangle)_{H}}{2}\right)^{-1}$ such that $\tau + \beta = \tau$. The temperature is then given by
\beq T = \frac{1}{\beta} = \frac{1}{2\pi}\frac{A_{H}(\partial_{r}\triangle)_{H}}{2} .\eeq
Because the imaginary time must be periodic, the Euclidean section method is not valid for dynamical spacetimes. Applying the same technique to a static conformal transformation ($\partial_{\tau}W =0$) of a static metric of the form
\beq \d s^{2} = WA^2 \triangle \d \tau^{2} + W\frac{\d r^{2}}{\triangle} + Wr^{2}\d\Omega^{2} \eeq
by a replacement $x= \alpha\sqrt{\triangle/W}$ we find $\alpha$ must take the value of $2W_{H}/\partial_{r}\triangle_{H}$ in order to avoid a conical singularity. Therefore the period of $\tau$ becomes $2\pi\alpha/(A_{H}W_{H})$ and the new temperature, $\tilde{T}$, is equal to the old temperature,
\beq \tilde{T} = \frac{1}{\tilde{\beta}} = \frac{1}{2\pi}\frac{A_{H}(\partial_{r}\triangle)_{H}}{2} .\eeq
With this method the temperature is invariant for static conformal transformations, provided we evaluate its value at the same location, a Killing horizon.  This agrees with the results of \cite{Marques:2011uq}. The invariance holds even as a limit for static conformal transformations that do not satisfy $\theta_{l}=0$ at the Killing horizon. The location of the relevant horizon in this method is located by the origin of the polar coordinates. When the horizon is no longer a Killing horizon such as the static veiled Schwarzschild example (\ref{veiledSchw}) then the horizon fails to be the origin of polar coordinates.

On a purely formal level the condition of vanishing conical deficit can also be applied at any other $r$, that is not the origin of polar coordinates, but in that case one will not get the canonical Hawking temperature for the Schwarzschild black hole. The Euclidean section method supports the argument that the Hawking temperature should be conformally invariant for Killing horizons of static spacetimes, even when the Killing horizon is not a $\theta_{l}=0$ surface.

\section{Gravitational anomaly}
%***********************************************************************************************************

Chiral theories in two dimensions have a gravitational anomaly \cite{AlvarezGaume:1983ig}. For a scalar field it takes the form \cite{Bertlmann:2000da}
\beq \nabla_{i}T^{i}_{j} = \frac{\varepsilon^{kl}}{96\pi\sqrt{-g}}\partial_{l}\partial_{m}\Gamma^{m}_{jk} .\eeq
In four-dimensional spherically symmetric metrics we can expand a test scalar field with arbitrary self-interactions in terms of partial wave modes and take a near horizon limit. In this limit the action for each partial scalar waves reduces to a free massless scalar field in two dimensions (see \cite{Robinson:2005pd} and references therein for further details). If we eliminate either the out-going or the in-going modes this becomes a two-dimensional chiral scalar field. Since the two-dimensional chiral theory is anomalous, a flux must occur to cancel the anomaly and restore general covariance. The flux, $\Phi$, can be derived by examining energy-momentum terms on either side of the horizon from whence it is found
\beq \Phi = \frac{\varepsilon^{ir}}{96\pi}\partial_{j}\Gamma^{j}_{ti} ,\eeq
where $t$ and $r$ are coordinates in the two-dimensional space normal to a horizon slicing and the right hand side is evaluated on the horizon. This formula does not define a tensor and in dynamical situations the result will depend on the choice of $t$ coordinate. The flux per particle mode is related to the temperature of a beam of massless blackbody radiation in the four-dimensional theory by
\beq \label{Phiflux} \Phi = \frac{\pi}{12}T^{2} ,\eeq
which provides a means of calculating the horizon temperature, $T_{H}$. This method was applied to the dynamical Vaidya spacetime in \cite{Vagenas:2006qb}, with the advanced null time $v$ chosen as the $t$ coordinate. For a general spherically symmetric metric (\ref{conftransEF}) with $W=1$, we find
\beq \Phi = \frac{1}{96\pi}\left[ \frac{\ddot{A}}{A} - \left(\frac{\dot{A}}{A}\right)^{2} + \frac{A'\dot{\triangle}}{2} + \frac{\dot{A}\triangle '}{2} + \frac{(A\triangle')^{2}}{2}\right] \eeq
when evaluated on the trapping horizon $\triangle = 0$. In this formula dots denote partial derivate with respect to $v$ and dashes denote partial derivative with respect to $r$. For the restricted case $A=1$, this reduces to
\beq \Phi = \frac{1}{96\pi}\frac{(\triangle')^{2}}{2} ,\eeq
which gives a dynamical temperature that agrees with most expectations \cite{Nielsen:2007ac} \cite{Pielahn:2011ra}. 

In \cite{Vagenas:2006qb} the formalism was applied to a null surface (the event horizon) in the Vaidya spacetime. Applying it there led to a flux that did not satisfy (\ref{Phiflux}) and this was interpreted as a non-thermal flux due to the dynamical spacetime. Our result is slightly more general than theirs (our function $\triangle(v,r)$ can be any function of $v$ and $r$ rather than just the Vaidya form $\triangle(v,r) = 1-2m(v)/r$ and if we evaluate it on the surface $\triangle=0$, rather than a null surface, we find a temperature that more closely agrees with expectations. However, the effective field theory method of discarding the ingoing modes is less compelling when applied to the trapping horizon as it does not mark the boundary of where these modes become inaccessible. The applicability of this method also depends on the choice of time coordinate as other choices give different results.

For the general conformal transformed metric of the form (\ref{conftransEF}), evaluated at $\triangle =0$, we have instead
\bea \Phi & =  & \frac{1}{96\pi}\Bigg[\frac{\ddot{W}}{W} - \left(\frac{\dot{W}}{W}\right)^{2} + \frac{A\dot{\triangle}}{2}\frac{W'}{W} + \frac{A\triangle'}{2}\frac{\dot{W}}{W} \nonumber \\ & & + \frac{A^{2}}{2}\left(\triangle'\right)^{2} + \frac{A'\dot{\triangle}}{2}-\left(\frac{\dot{A}}{A}\right)^{2}+\frac{\ddot{A}}{A} + \triangle'\dot{A}\Bigg] .\eea
Only for a static conformal transformation of a static spacetime will this generally be conformally invariant at $\triangle = 0$. This includes the static example (\ref{veiledSchw}) when the flux is calculated in the limit $\triangle \rightarrow 0$ and applies even if $\triangle=0$ is not a Killing horizon.

By the arguments given in \cite{Flanagan:2004bz} and elsewhere, any dynamical conformal transformation of a static spacetime should leave the physics invariant. But, while the choice adopted in \cite{Vagenas:2006qb} of simply using the advanced null coordinate $v$ may be valid, this remains to be shown, so it is perhaps not surprising that the gravitational anomaly calculation only gives a conformally invariant temperature in the static limit.

\section{The Hamilton-Jacobi tunneling method}
%****************************************************************************************************************

The tunneling method proceeds by relating a pole in the path integral of solutions of the Hamilton-Jacobi equation to a transition amplitude across the horizon. This method is very relevant to our current investigation because it forms the basis of the claims in \cite{Di Criscienzo:2007fm} that the tunneling is associated with zero expansion quasi-local horizons of the trapping horizon form. Further details of the method can be found in \cite{Vanzo:2011wq} and references therein.

To illustrate the methodology we can examine the simple static veiled Schwarzschild example (\ref{veiledSchw}) considered in \cite{Deruelle:2010ht}. This is a specific example of the general \textit{synchronous gauge} class considered in \cite{hay0906}. The radial outgoing null vector has vanishing expansion at $r=3M$, where the radial ingoing null vector also has vanishing expansion. Using the eikonal/geometric optics approximation we can write a tunneling action, $I$, as
\beq \label{eikonal} I = \int \d x^{i}\partial_{i}I \equiv \int \omega \d t + \int k \d r .\eeq
Because the spacetime admits a Killing vector $\partial_{t}$, the factor $\omega$ is a constant along the motion provided $\partial_{a}I$ defines the dual tangent of an affinely parameterised geodesic, $(\partial^{a}I)\nabla_{a}(\partial_{b}I) = 0$. The Hamilton-Jacobi equation gives
\beq \label{HamJaceqn} g^{ab}\partial_{a}I\partial_{b}I = 0 .\eeq
This is just the null condition for trajectories with tangents $\partial_{a}I$ which expresses the null ray approximation of geometric optics. There is a pole at $r=2M$ since for the metric (\ref{veiledSchw})
\beq \partial_{r}I = \pm \frac{\partial_{t}I}{\triangle} ,\eeq
assuming that $\partial_{t}I$ is regular at $\triangle = 0$. The method proposed by Di Criscenzo et al. \cite{hay0906} can also be examined here. They define a scalar invariant $\chi$, that for the metric (\ref{veiledSchw}) is given by
\beq \chi \equiv \gamma^{ij}\partial_{i}R\partial_{j}R = \frac{\left(r-3M\right)^{2}}{r^{2}\triangle} .\eeq
It is singular in the limit $r\rightarrow 2M$ and zero at $r=3M$ which indicates the presence of a horizon at $r=3M$ \cite{hay0906}. The Kodama vector is
\beq K^{i} \equiv \frac{1}{\sqrt{-\gamma}}\epsilon^{ij}\partial_{j}R = \left(\frac{(r-3M)}{r\sqrt{\triangle}},0,0,0\right) .\eeq
Again it has singular components as $r\rightarrow 2M$ and vanishes at $r=3M$. The ``invariant energy'' is then
\beq \omega_{K} \equiv -K^{i}\partial_{i}I = \frac{(r-3M)}{r\sqrt{\triangle}}\partial_{t}I .\eeq
The ``invariant'' energy vanishes at $r=3M$ if $\partial_{t}I = \omega$ is regular there, which it will be as it is a constant along the motion. In terms of the ``invariant energy'' the Hamilton-Jacobi equation gives
\beq \partial_{r}I = \pm\frac{r}{\sqrt{\triangle}}\frac{\omega_{K}}{(r-3M)} .\eeq
This gives the impression of a pole at $r=3M$ where the expansion is vanishing, but there is no pole here because $\omega_{K}$ is also vanishing linearly at $r=3M$ when $\omega=\partial_{t}I$ is regular. The statement after equation (2.29) of \cite{hay0906} that the expression for $\partial_{r}I$ \textit{``has a pole at the horizon''} is not true if in this case the trapping horizon is meant, because the ``invariant'' energy is zero there when $\partial_{t}I$ is regular. There is only a pole in the solution of the Hamilton-Jacobi equation at $\triangle=0$ which is not at $\chi=0$.

For the time dependent conformal transformation of Schwarzschild (\ref{conftransEF}), that contains true future outer trapping horizons not located at $\triangle =0$, once again the Hamilton-Jacobi equation gives a pole for $\partial_{r}I$ at $\triangle =0$, provided $\partial_{t}I \neq 0$ at those points. This can be seen directly from the Hamilton-Jacobi equation (\ref{HamJaceqn}) that is manifestly conformally invariant, as it should be. We have assumed that $\omega=\partial_{t}I$ is a constant of the motion, which is certainly true  if $\partial_{a}I$ is the dual tangent of an affinely parameterised geodesic and $k^{a}$ is a timelike Killing vector such that $\omega = k^{a}\partial_{a}I$, but this remains true under a conformal transformation since $\omega$ is a constant along the worldline which remains unchanged. This can be seen explicitly since
\beq \partial^{a}I\nabla_{a}\left(k^{b}\partial_{b}I\right) = \left(\partial^{a}I\right)\left(\partial^{b}I\right)\nabla_{a}k_{b} +k^{b}\left(\partial^{a}I\right)\nabla_{a}\partial_{b}I .\eeq
The first term on the right hand side vanishes for a conformal Killing vector field that satisfies $\nabla_{a}k_{b}+\nabla_{b}k_{a}=g_{ab}k^{c}\partial_{c}W$ and the second term vanishes because an affine null geodesic $\partial_{a}I$ remains an affine geodesic under a conformal transformation if we assume that $\partial_{a}I \rightarrow \partial_{a}I$. Thus we find there is no pole at $r=3M$ even for the time dependent conformal transformation (\ref{timedepmetric}) which has a future outer trapping horizon at $r=3M$. The norm of the Kodama vector also vanishes at $r=3M$ where $\theta_{l}=0$, not at $\triangle = 0$. 

%********************************************************************************************
\subsection{Hamilton-Jacobi equation and the eikonal approximation }
%********************************************************************************************

An important assumption for the tunneling method above is the validity of the eikonal approximation and an important condition for the validity of the eikonal approximation is the small wavelength limit \cite{Landau:1974}. Under a conformal transformation that changes spacetime lengths, one may wonder whether this condition is still fulfilled. For Schwarzschild black holes, asymptotic observers see a typical radiation wavelength that is of the order of the areal radius of the black hole. In the near horizon limit of the tunneling method, the wavelength of this radiation will be small relative to the size of the black hole due to significant redshifting. 

A conformal transformation can alter the geometrical redshift along a particle's path. For example, a radially emitted photon in the metric of (\ref{veiledSchw}) will undergo no geometrical redshifting when traversing from near the black hole to an observer far away. The wavelength of the Hawking radiation observed far away will still be of order $M$, $\lambda_{\infty} \sim M$. But in this case, because of the conformal factor, the size of the black hole is much larger than $M$ (formally infinity in the example) and hence the eikonal approximation will still be valid in the near horizon region.

A related condition for the validity of the eikonal approximation is that the geometry should be changing slowly with respect to the typical frequencies of the radiation. The Schwarzschild exterior is explicitly static, but it is expected that Hawking radiation is also emitted in dynamical situations. For the eikonal approximation to break down for black holes requires the horizon to be growing very radidly, or shrinking very rapidly, far faster than is expected by Hawking radiation in all but the very end of the evaporation process when semi-classical approximations are likely to be invalid anyway.

Very simple examples of horizons are also given by the cosmological horizons of the Friedmann-Robertson-Walker (FRW) spacetimes. These horizons are observer dependent in the sense that they bound the region which can send future-directed signals to a given observer and for different observer locations the horizons have different locations. Horizons in FRW spacetimes with matter are explicitly time dependent through the time dependence of the Hubble parameter, $H$, and for models that are matter dominated one can show that the corresponding cosmological horizons are not infinite redshift surfaces. Consider the FRW metric in standard comoving isotropic coordinates,
\beq
\label{FRW} \d s^2=-\d t^{2}+a(t)^{2}\left(\frac{\d r^2}{1-kr^2}+r^2 \d\Omega^2\right) .
\eeq
The horizons in FRW are Past Inner Trapping Horizons if the deceleration parameter $q\equiv -\frac{\ddot{a}a}{\dot{a}^{2}}$ satisfies $q<1$. Examples that satisfies this condition are pure de Sitter space or the $\Lambda$CDM model of the current epoch where both outgoing and ingoing null rays are forced to move outwards beyond the horizon. If an emitter with four-velocity $u^{\mu}_{e}$, sends a light ray to an observer along a null path with tangent $n^{\mu}$, the relative light redshift, $z$, that is calculated by an observer with four-velocity $u^{\mu}_{o}$ will be,
\beq 1+z=\frac{(n_\mu u^\mu)_{e}}{(n_\mu u^\mu)_o}. \eeq
Under a conformal transformation this will remain unchanged since both $n_{\mu}$ and $u^{\mu}$ remain unchanged. A comoving observer's four-velocity is $u^{\mu}=(1,0,0,0)$, so, in the standard FRW metric (\ref{FRW}) the redshift is just $1+z=n^{t}_{e}/n^{t}_{o}$.
Solving the affine geodesic equation, $n^{\mu}\bigtriangledown_\mu n^{\nu}=0$, for a radial null geodesic $\left( n^{t}\right) ^{2}=\frac{a(t)^2}{1-kr^2}\left( n^{r}\right) ^2$ gives,
\beq \frac{\d n^t}{\d\lambda}=-\Gamma_{\mu\nu}^t n^\mu n^\nu=-\Gamma_{rr}^t \frac{1-kr^2}{a(t)^2} (n^t)^2 , \eeq
where $\Gamma_{rr}^t=\frac{\dot{a}(t)a(t)}{1-kr^{2}}$. Thus the solution is $n^{t}=C/a$ for $C$ some integration constant, and we obtain $1+z = \frac{a_{o}}{a_{e}}$ as is standard.

For a photon emitted in the vicinity of an observer-dependent horizon located at $r=1/\dot{a}$, for a comoving observer located at $r=0$, we can calculate the measured redshift of the photon after it travels from the horizon to the observer at $r=0$. The measured redshift $z$ is not generally infinite for a photon emitted a small finite distance from the cosmological horizon. An infinite redshift is only obtained in the limit that the cosmological horizon is the event horizon of the observer. This occurs for pure de Sitter space, but not in more general cases such as a $\Lambda=0$ matter dominated universe, where the scale factor behaves as $a\sim t^{2/3}$. In this case the relative redshift will be just $1+z=9/4$. Therefore, we cannot rely on the geometric optics approximation for all apparent $\theta_{n}=0$ cosmological horizons. This condition is not imposed in \cite{Cai:2008gw} that otherwise discusses the behaviour of radiation from cosmological apparent horizons.

In \cite{Visser:2001kq} the condition given for the validity of the eikonal split (\ref{eikonal}) is
\beq \omega \gg max\left(\left|\frac{\dot{c}}{c}\right|,\left|\frac{\dot{v}}{v}\right|\right) ,\eeq
where $v$ and $c$ are metric functions in Painlev\'{e}-Gullstrand coordinates.
\beq \d s^{2} = -(c^{2}-v^{2})\d t^{2} +2v\d t\d r + \d r^{2} + r^{2}\d\Omega^{2} \eeq
This is the condition that the geometry not be changing rapidly with respect to the typical frequency of the radiation, $\omega$. In the Schwarzschild spacetime this is the typical frequency of the radiation measured by inertial observers at infinity, but it is also a lower bound on the frequency measured by any observer anywhere outside the horizon. One advantage of using Painlev\'{e}-Gullstrand coordinates is that the time coordinate coincides with the $k=0$ FRW comoving time, the corrdinate $t$ of (\ref{FRW}). For the FRW spacetimes considered in \cite{Cai:2008gw} with $k=0$, we can transform equation (\ref{FRW}) by changing the radial coordinate to $\tilde{r}=ar$, in which case the metric is put in Painlev\'{e}-Gullstrand form with $c=1$ and
\beq v = \tilde{r}H .\eeq
Therefore we have
\beq \frac{\dot{v}}{v} = \frac{\dot{H}}{H} = \frac{\ddot{a}}{\dot{a}}-\frac{\dot{a}}{a} .\eeq
and
\beq \omega \sim |\kappa| = H = \frac{\dot{a}}{a} .\eeq
Here we impose an absolute magnitude to obtain positive frequencies and temperatures. To satisfy the eikonal condition we therefore need
\beq \frac{\ddot{a}}{\dot{a}} \sim \frac{\dot{a}}{a} .\eeq
From \cite{Faraoni:2011zy} we have that the norm of the apparent horizon in FRW spacetime is
\beq N^{a}N_{a} = \frac{1}{a^{2}}\left(1 - \left(\frac{\ddot{a}a}{\dot{a}^{2}}\right)^{2}\right) \eeq
and so the horizon needs to be ``almost null'' for the eikonal approximation to be applicable. This is not satisfied for all apparent horizons in FRW spacetimes. From the Friedmann equations for an FRW with $k=0$ and $\Lambda=0$ and perfect fluid equation of state $p=w\rho$, we find
\beq \frac{\ddot{a}a}{\dot{a}^{2}} = -\frac{1}{2}\left(1+3w\right), \eeq
which needs $w\sim-1$ or $w\sim 1/3$ for the eikonal split (\ref{eikonal}), which are cosmological constant dominated and radiation dominated respectively. In \cite{Cai:2008gw} it is mentioned that an important requirement in deriving the Friedmann equations from the first law for the horizons is $\dot{\tilde{r}}_{A}=0$, that the location of the apparent horizon is fixed in an infinitesimal time interval. This is only works if the horizon is close to being null (slowly evolving) and for the case $k=0$, $\Lambda=0$ this requires the universe to be dark energy dominated $w\sim -1$.
%

%****************************************************************************************************
\section{Bogoliubov transformations}
%****************************************************************************************************

The original demonstration of the Hawking effect calculated the Bogoliubov coefficients relating the field components at future null infinity and the field components at past infinity \cite{Hawking:1974sw}. It was argued in \cite{Jacobson:1993pf} that it is trivial to see that this argument is conformally invariant for a free, conformally coupled field. In \cite{Marques:2011uq} this was demonstrated explicitly for a static, Killing horizon preserving, conformal transformation. To see how this works we follow the method of \cite{Ford:1997hb} and consider a thin hollow shell collapsing in a spherical spacetime. The thin shell provides calculational convenience and does not entail a critical loss of generality.

The field at future null infinity can be decomposed into positive frequency radial modes using a radial null coordinate $u$. The field at past null infinity can be decomposed into positive frequency radial modes using a null coordinate $v$. Outgoing radial modes that reach future null infinity can be traced back through the spacetime to past null infinity, where they will be incoming modes. These incoming modes from past null infinity will cross the infalling thin shell before entering the flat region inside the shell. Then they will reach $r=0$ the centre of spherical symmetry, before exiting the flat region, crossing the thin shell again and escaping. To relate fields at past and future infinity one calculates the Bogoliubov coefficients. The Bogoliubov coefficients will be non-trivial, implying particle production or population of modes, if there is a non-linear relation between the null coordinates $u$ and $v$ following the procedure of \cite{Ford:1997hb} or \cite{Barcelo:2010xk}.

To illustrate the method consider the exterior of the thin shell to be pure Schwarzschild. Inside the hollow shell the metric is flat
\beq \d s^{2} = -\d T^{2} + \d R^{2} + R^{2}\d \Omega^{2} .\eeq
Outside the shell the metric is Schwarzschild
\beq \d s^{2} = -\triangle\d t^{2} + \frac{\d r^{2}}{\triangle} + r^{2}\d\Omega^{2} .\eeq
There are three points of interests to find the relation between $u$ and $v$. At the first crossing of the thin shell we will have
\beq V_{cross} = av_{cross} + b ,\eeq
where $V$ is the advanced radial null coordinate inside the flat region and $a$ and $b$ are constants. The rays then follow constant $v$ paths until they hit $R=0$. At $R=0$ in the flat region we can impose the simple relation
\beq V = U \eeq
between the advanced coordinate of incoming rays and their corresponding outgoing retarded radial null coordinate, $U$. At the second crossing of the thin shell, for paths that run close to the event horizon, we will have
\beq \label{ucross} u_{cross} \simeq -4M\ln\left(\frac{T_{_{EH}}-T_{cross}}{B}\right) ,\eeq
where $T_{_{EH}}$ is the time at which the last escaping null ray crosses the thin shell in the flat space coordinates and $B$ is a constant. This shell crossing time is at a finite value in the flat space. This last equation arises because at the thin shell the metrics must match each other and the radial coordinates can be equated, $r_{cross} = R_{cross}$, so one obtains
\beq \label{metricmatch} -1 + \left(\frac{\d R}{\d T}\right)^{2} = \triangle\left(\frac{\d t}{\d T}\right)^{2} - \frac{1}{\triangle}\left(\frac{\d R}{\d T}\right)^{2} .\eeq
An outgoing ray in the flat region will meet the thin shell at
\beq \label{Rcross} R_{cross} = 2M + (T_{0}-T_{cross}) .\eeq
The outgoing null coordinate in the Schwarzschild region can be written as $u=t-r_{*}$ where $r_{*}$ is the tortoise coordinate
\beq \label{rstar} r_{*} = r + 2M\ln\left(\frac{r-2M}{2M}\right) .\eeq
Integrating the condition (\ref{metricmatch}) with the condition (\ref{Rcross}) and (\ref{rstar})  gives (\ref{ucross}). The relations from the three points can now be combined to give the relation between the null coordinates at past and future infinity
\beq u = -4M\ln\left(\frac{v_{0}-v}{C}\right) ,\eeq
where $C$ is a constant and $v_{0}$ refers to the value of $v$ which exactly meets the event horizon.

This method is conformally invariant because equations (\ref{metricmatch}) and (\ref{Rcross}) are manifestly conformally invariant and also the tortoise coordinate, $r_{*}$, is manifestly invariant \cite{Marques:2011uq} since
\beq \d r_{*} = \sqrt{\frac{g_{rr}}{-g_{tt}}}\d r .\eeq
The coordinates do not change, and neither do the null paths, so it is not particularly surprising that the relation between $u$ and $v$ is unchanged. The relevant temperature remains the same, $1/4M$, as noted in \cite{Jacobson:1993pf}. The relevant horizon is not directly identified, except through the logarithmic pile-up of (\ref{ucross}), described as peeling in \cite{Barcelo:2010pj}.

%******************************************************************************************************
\section{Conclusions and discussion}
%******************************************************************************************************

For radial trajectories in spherically symmetric spacetimes, as considered here, the geometry is effectively 1+1 dimensional. The 1+1 dimensional Schwarzschild spacetime is conformally flat. This means that for the 1+1 dimensional Schwarzschild spacetime there are transformations for which one would not expect any non-linear relation between asymptotic null coordinates. But the four dimensional Schwarzschild solution is not conformally flat, as can be seen by the non-zero components of the conformally invariant Weyl tensor - which in the vacuum, $R_{ab}=0$, case are equivalent to the components of the Riemann curvature tensor.

We have also focussed on rather simple conformal transformations for simplicity and ease of calculation. There is obviously a much larger class of conformal transformations one could consider. Our work here has focussed on the spherically symmetric Schwarzschild spacetime, but it is likely that the results can be generalised to less symmetric situations.

It may not ultimately be resolvable whether a particular surface is emitting Hawking radiation in an actual experimental set-up. But an important question is what are the conditions necessary for the production of Hawking radiation from gravitational fields? If these minimal conditions were met then one would expect Hawking radiation to be produced, leading to a flux of energy and possible backreaction effects. Without these minimal conditions there would be no Hawking radiation. We have argued that conformal transformations, although a purely formal mathematical operation, can provide valuable insight into this question.

There are good reasons to believe that conformal transformations should not change semi-classical experimental predictions. A strong version of the physical equaivalence of conformal frames would be that all entitites entering physical theories should be defined in a conformally invariant manner. The location of a zero expansion surface such as used in trapping horizons is not invariant under a conformal transformation.  We have attempted to partially address the issue of the origin of Hawking radiation by examining a number of semi-classical effective derivations for the Hawking effect. Each of these has its own strengths and weaknesses and our confidence in their reliability is often dependent on the extent to which they give similar answers. 

The Euclidean section method only works for static configurations and only for configurations that preserve the Killing horizon, but gives rise to a conformally invariant result for the imaginary time period and hence temperature. The gravitational anomaly calculation works for static configurations even without a Killing horizon and in this case gives a conformally invariant flux. It is not conformally invariant when applied to dynamical cases as in \cite{Vagenas:2006qb}, although there are reasons to distrust a simple extension in the dynamical case. 

The Hamilton-Jacobi tunneling method appears to work for dynamical spacetimes and has been applied to such for both black holes \cite{hay0906} and cosmological spacetimes \cite{Cai:2008gw}. In the dynamical case care must be taken that the conditions for the eikonal approximation are satisfied. The Hamilton-Jacobi equation is manifestly conformally invariant and we have shown how standard treatments can lead to a conformally invariant result. The Bogoliubov coefficient method works for a collapsing spacetime by considering null rays close to the causal horizon. Based as it is on the paths of null rays, it is perhaps the method that is most explicitly conformally invariant. This is true even for propagation near to conformal Killing horizons in a time dependent spacetime. The method has been criticised for relying on rays that pass closest to the horizon and are thus trans-Planckian, but this method also does not actually require a horizon since it is strictly limited to the domain outside the horizon.

Other authors have come to the conclusion that no horizon at all is strictly necessary for the Hawking effect \cite{Barcelo:2006uw,Barcelo:2010pj,Barcelo:2010xk} and horizons may not even form \cite{Barbado:2011ai}. It may well be that all that is required is a curved spacetime. Dense compact horizonless objects may emit very faint Hawking radiation - a flux of radiation away from the source - but at a rate that depends on how massive the object is and how close it is to forming a horizon. If this is indeed true it would imply that the Hamilton-Jacobi tunneling picture is not an altogether profitable picture, as it clearly relies on some type of horizon to be operable. The Bogoliubov coefficient method however, is applicable even when no horizon forms \cite{Barcelo:2010pj} and remains conformally invariant in agreement with expectations.

The reader may be left with the impression that the problems raised in this work will only affect non-Einstein theories or situations where standard units are allowed to vary in space and time. But this is not correct. Firstly, the rather simple and rather elegant approach proposed in \cite{Jacobson:1993vj}, for dealing with non Einstein gravity theories will not be strictly applicable for vanishing expansion horizons such as the trapping horizon. This drawback though may be partly alleviated by a modified quasi-local horizon definition based on the Wald entropy \cite{Faraoni:2011zy}. This definition is conformally invariant and will allow one to map to the Einstein conformal frame whilst still treating the same physical surface. This may be useful for extending numerical codes to non-Einstein gravity models, by mapping equations to the Einstein frame.

But, more challengingly, the metrics obtained by simple conformal transformations of the Schwarzschild solution, such as (\ref{veiledSchw}), could in principle occur as the metric solutions of the Einstein equations in the Einstein conformal frame, albeit for rather strange matter fields that violate various energy conditions. In these cases the proposal in \cite{Faraoni:2011zy} would not locate the horizon at $\triangle = 0$, since the Wald entropy would be just one quarter of the area and we would be led to a static, timelike, two-way traversable horizon at $\theta_{l}=0$, not the null surface $\triangle =0$. This observation in itself is however not enough to show that null horizons should be preferred at all times, since there are many situations where this is questionable too.

\section{Acknowledgments}

A.B.N. is very grateful for generous support from the Alexander von Humboldt foundation and J.T.F. is grateful to the Max Planck Institute for Gravitational Physics in Potsdam for hospitality during April to June 2011, where the majority of this work was completed. The authors are grateful to Carlos Barcelo for useful discussions.


\bigskip


\begin{thebibliography}{999}

%\cite{Gibbons:1977mu}
\bibitem{Gibbons:1977mu}
  G.~W.~Gibbons, S.~W.~Hawking,
  %``Cosmological Event Horizons, Thermodynamics, and Particle Creation,''
  Phys.\ Rev.\  {\bf D15 } (1977)  2738-2751.

%\cite{Hawking:1974sw}
\bibitem{Hawking:1974sw}
  S.~W.~Hawking,
  %``Particle Creation by Black Holes,''
  Commun.\ Math.\ Phys.\  {\bf 43 } (1975)  199-220.

%\cite{Christensen:1977jc}
\bibitem{Christensen:1977jc}
  S.~M.~Christensen and S.~A.~Fulling,
  %``Trace Anomalies and the Hawking Effect,''
  Phys.\ Rev.\ D {\bf 15} (1977) 2088.
  %%CITATION = PHRVA,D15,2088;%%

%\cite{Robinson:2005pd}
\bibitem{Robinson:2005pd}
  S.~P.~Robinson and F.~Wilczek,
  %``A Relationship between Hawking radiation and gravitational anomalies,''
  Phys.\ Rev.\ Lett.\  {\bf 95} (2005) 011303
  [gr-qc/0502074].
  %%CITATION = GR-QC/0502074;%%

%\cite{Parikh:1999mf}
\bibitem{Parikh:1999mf}
  M.~K.~Parikh, F.~Wilczek,
  %``Hawking radiation as tunneling,''
  Phys.\ Rev.\ Lett.\  {\bf 85 } (2000)  5042-5045.
  [arXiv:hep-th/9907001 [hep-th]].

%\cite{Vanzo:2011wq}
\bibitem{Vanzo:2011wq}
  L.~Vanzo, G.~Acquaviva and R.~Di Criscienzo,
  %``Tunnelling Methods and Hawking's radiation: achievements and prospects,''
  Class.\ Quant.\ Grav.\  {\bf 28} (2011) 183001
  [arXiv:1106.4153 [gr-qc]].
  %%CITATION = ARXIV:1106.4153;%%

%\cite{Di Criscienzo:2007fm}
\bibitem{Di Criscienzo:2007fm}
  R.~Di Criscienzo, M.~Nadalini, L.~Vanzo, S.~Zerbini, G.~Zoccatelli,
  %``On the Hawking radiation as tunneling for a class of dynamical black holes,''
  Phys.\ Lett.\  {\bf B657 } (2007)  107-111.
  [arXiv:0707.4425 [hep-th]].

%\cite{Hayward:2008jq}
\bibitem{Hayward:2008jq}
  S.~A.~Hayward, R.~Di Criscienzo, L.~Vanzo, M.~Nadalini, S.~Zerbini,
  %``Local Hawking temperature for dynamical black holes,''
  Class.\ Quant.\ Grav.\  {\bf 26 } (2009)  062001.
  [arXiv:0806.0014 [gr-qc]].

%\cite{Nielsen:2008cr}
\bibitem{Nielsen:2008cr}
  A.~B.~Nielsen,
  %``Black holes and black hole thermodynamics without event horizons,''
  Gen.\ Rel.\ Grav.\  {\bf 41 } (2009)  1539-1584.
  [arXiv:0809.3850 [hep-th]].

%\cite{Cai:2008gw}
\bibitem{Cai:2008gw}
  R.~-G.~Cai, L.~-M.~Cao, Y.~-P.~Hu,
  %``Hawking Radiation of Apparent Horizon in a FRW Universe,''
  Class.\ Quant.\ Grav.\  {\bf 26 } (2009)  155018.
  [arXiv:0809.1554 [hep-th]].

%\cite{Jacobson:1993pf}
\bibitem{Jacobson:1993pf}
  T.~Jacobson and G.~Kang,
  %``Conformal invariance of black hole temperature,''
  Class.\ Quant.\ Grav.\  {\bf 10} (1993) L201
  [gr-qc/9307002].
  %%CITATION = GR-QC/9307002;%%

%\cite{Koga:1998un}
\bibitem{Koga:1998un}
  J.~-i.~Koga, K.~-i.~Maeda,
  %``Equivalence of black hole thermodynamics between a generalized theory of gravity and the Einstein theory,''
  Phys.\ Rev.\  {\bf D58 } (1998)  064020.
  [gr-qc/9803086].

%\cite{Jacobson:1993vj}
\bibitem{Jacobson:1993vj}
  T.~Jacobson, G.~Kang and R.~C.~Myers,
  %``On black hole entropy,''
  Phys.\ Rev.\ D {\bf 49} (1994) 6587
  [gr-qc/9312023].
  %%CITATION = GR-QC/9312023;%%

%\cite{Nielsen:2010gm}
\bibitem{Nielsen:2010gm}
  A.~B.~Nielsen,
  %``The spatial relation between the event horizon and trapping horizon,''
  Class.\ Quant.\ Grav.\  {\bf 27} (2010) 245016
  [arXiv:1006.2448 [gr-qc]].
  %%CITATION = CQGRD,27,245016;%%

%\cite{Faraoni:2011zy}
\bibitem{Faraoni:2011zy}
  V.~Faraoni, A.~B.~Nielsen,
  %``The horizon-entropy increase law for causal and quasi-local horizons and conformal field redefinitions,''
  Class.\ Quant.\ Grav.\  {\bf 28}, 175008 (2011) .
  [arXiv:1103.2089 [gr-qc]].

%\cite{Scheel:1994yn}
\bibitem{Scheel:1994yn}
  M.~A.~Scheel, S.~L.~Shapiro and S.~A.~Teukolsky,
  %``Collapse to black holes in Brans-Dicke theory. 2. Comparison with general relativity,''
  Phys.\ Rev.\ D {\bf 51} (1995) 4236
  [gr-qc/9411026].
  %%CITATION = GR-QC/9411026;%%

%\cite{Flanagan:2004bz}
\bibitem{Flanagan:2004bz}
  E.~E.~Flanagan,
  %``The Conformal frame freedom in theories of gravitation,''
  Class.\ Quant.\ Grav.\  {\bf 21 } (2004)  3817.
  [gr-qc/0403063].

%\cite{Faraoni:2006fx}
\bibitem{Faraoni:2006fx}
  V.~Faraoni, S.~Nadeau,
  %``The (pseudo)issue of the conformal frame revisited,''
  Phys.\ Rev.\  {\bf D75 } (2007)  023501.
  [gr-qc/0612075].

%\cite{Deruelle:2010ht}
\bibitem{Deruelle:2010ht}
  N.~Deruelle, M.~Sasaki,
  %``Conformal equivalence in classical gravity: the example of 'veiled' General Relativity,''
  [arXiv:1007.3563 [gr-qc]].

%\cite{Dicke:1961gz}
\bibitem{Dicke:1961gz}
  R.~H.~Dicke,
  %``Mach's principle and invariance under transformation of units,''
  Phys.\ Rev.\  {\bf 125 } (1962)  2163-2167.

%\cite{Visser:2009pw}
\bibitem{Visser:2009pw}
  M.~Visser, C.~Barcelo, S.~Liberati and S.~Sonego,
  %``Small, dark, and heavy: But is it a black hole?,''
  arXiv:0902.0346 [gr-qc].
  %%CITATION = ARXIV:0902.0346;%%

%\cite{Marques:2011uq}
\bibitem{Marques:2011uq}
  G.~T.~Marques and M.~E.~Rodrigues,
  %``Equivalence of the Hawking temperature in conformal frames,''
  Eur.\ Phys.\ J.\ C {\bf 72} (2012) 1891
  [arXiv:1110.0079 [gr-qc]].
  %%CITATION = ARXIV:1110.0079;%%

%\cite{Nielsen:2010wq}
\bibitem{Nielsen:2010wq}
  A.~B.~Nielsen, M.~Jasiulek, B.~Krishnan and E.~Schnetter,
  %``The slicing dependence of non-spherically symmetric quasi-local horizons in
  %Vaidya Spacetimes,''
  arXiv:1007.2990 [gr-qc].
  %%CITATION = ARXIV:1007.2990;%%

%\cite{hay0906}
\bibitem{hay0906}
R.~Di Criscienzo, S.~A.~Hayward, M.~Nadalini  L.~Vanzo and
S.~Zerbini,
  %``Hamilton-Jacobi Tunneling Method for Dynamical Horizons in Different Coordinate Gauges,''
  Class.\ Quant.\ Grav.\  {\bf 27}, 015006 (2010) .

%\cite{Magnano:1993bd}
\bibitem{Magnano:1993bd}
  G.~Magnano, L.~M.~Sokolowski,
  %``On physical equivalence between nonlinear gravity theories and a general relativistic selfgravitating scalar field,''
  Phys.\ Rev.\  {\bf D50 } (1994)  5039-5059.
  [gr-qc/9312008].

%\cite{Ashtekar:2003jh}
\bibitem{Ashtekar:2003jh}
  A.~Ashtekar, A.~Corichi and D.~Sudarsky,
  %``Nonminimally coupled scalar fields and isolated horizons,''
  Class.\ Quant.\ Grav.\  {\bf 20} (2003) 3413
  [gr-qc/0305044].
  %%CITATION = GR-QC/0305044;%%

%\cite{Hayward:1993wb}
\bibitem{Hayward:1993wb}
  S.~A.~Hayward,
  %``General laws of black hole dynamics,''
  Phys.\ Rev.\ D {\bf 49} (1994) 6467.
  %%CITATION = PHRVA,D49,6467;%%

%\cite{Wald:1993nt}
\bibitem{Wald:1993nt}
  R.~M.~Wald,
  %``Black hole entropy is the Noether charge,''
  Phys.\ Rev.\ D {\bf 48} (1993) 3427
  [gr-qc/9307038].
  %%CITATION = GR-QC/9307038;%%

%\cite{Misner:1964je}
\bibitem{Misner:1964je}
  C.~W.~Misner and D.~H.~Sharp,
  %``Relativistic equations for adiabatic, spherically symmetric gravitational collapse,''
  Phys.\ Rev.\  {\bf 136} (1964) B571.
  %%CITATION = PHRVA,136,B571;%%

%\cite{DiCriscienzo:2010zza}
\bibitem{DiCriscienzo:2010zza}
  R.~Di Criscienzo, S.~A.~Hayward, M.~Nadalini, L.~Vanzo and S.~Zerbini,
  %``Hamilton-Jacobi tunneling method for dynamical horizons in different coordinate gauges,''
  Class.\ Quant.\ Grav.\  {\bf 27} (2010) 015006.
  %%CITATION = CQGRD,27,015006;%%

%\cite{Hayward:1997jp}
\bibitem{Hayward:1997jp}
S.~A.~Hayward,
  %``Unified first law of black-hole dynamics and relativistic
 %thermodynamics,''
  Class.\ Quant.\ Grav.\  {\bf 15}, 3147 (1998).

%\cite{AlvarezGaume:1983ig}
\bibitem{AlvarezGaume:1983ig}
  L.~Alvarez-Gaume and E.~Witten,
  %``Gravitational Anomalies,''
  Nucl.\ Phys.\ B {\bf 234} (1984) 269.
  %%CITATION = NUPHA,B234,269;%%

%\cite{Bertlmann:2000da}
\bibitem{Bertlmann:2000da}
  R.~A.~Bertlmann and E.~Kohlprath,
  %``Two-dimensional gravitational anomalies, Schwinger terms and dispersion relations,''
  Annals Phys.\  {\bf 288} (2001) 137
  [hep-th/0011067].
  %%CITATION = HEP-TH/0011067;%%

%\cite{Vagenas:2006qb}
\bibitem{Vagenas:2006qb}
  E.~C.~Vagenas and S.~Das,
  %``Gravitational anomalies, Hawking radiation, and spherically symmetric black holes,''
  JHEP {\bf 0610} (2006) 025
  [hep-th/0606077].
  %%CITATION = HEP-TH/0606077;%%

%\cite{Nielsen:2007ac}
\bibitem{Nielsen:2007ac}
  A.~B.~Nielsen and J.~H.~Yoon,
  %``Dynamical surface gravity,''
  Class.\ Quant.\ Grav.\  {\bf 25} (2008) 085010
  [arXiv:0711.1445 [gr-qc]].
  %%CITATION = ARXIV:0711.1445;%%

%\cite{Pielahn:2011ra}
\bibitem{Pielahn:2011ra}
  M.~Pielahn, G.~Kunstatter and A.~B.~Nielsen,
  %``Dynamical Surface Gravity in Spherically Symmetric Black Hole Formation,''
  Phys.\ Rev.\ D {\bf 84} (2011) 104008
  [arXiv:1103.0750 [gr-qc]].
  %%CITATION = ARXIV:1103.0750;%%

%\cite{Landau:1974}
\bibitem{Landau:1974}
  L.~D.~Landau, E.~M.~Lifshitz,,
  ``The Classical Theory of Fields, 4th Ed.''
  Butterworth-Heinemann, Oxford, (1974)

%\cite{Visser:2001kq}
\bibitem{Visser:2001kq}
  M.~Visser,
  %``Essential and inessential features of Hawking radiation,''
  Int.\ J.\ Mod.\ Phys.\  {\bf D12 } (2003)  649-661.
  [hep-th/0106111].

%\cite{Ford:1997hb}
\bibitem{Ford:1997hb}
  L.~H.~Ford,
  %``Quantum field theory in curved space-time,''
  In *Campos do Jordao 1997, Particles and fields* 345-388
  [gr-qc/9707062].
  %%CITATION = GR-QC/9707062;%%

%\cite{Barcelo:2010xk}
\bibitem{Barcelo:2010xk}
  C.~Barcelo, S.~Liberati, S.~Sonego, M.~Visser,
  %``Hawking-like radiation from evolving black holes and compact horizonless objects,''
  JHEP {\bf 1102 } (2011)  003.
  [arXiv:1011.5911 [gr-qc]].

%\cite{Barcelo:2010pj}
\bibitem{Barcelo:2010pj}
  C.~Barcelo, S.~Liberati, S.~Sonego, M.~Visser,
  %``Minimal conditions for the existence of a Hawking-like flux,''
  Phys.\ Rev.\  {\bf D83 } (2011)  041501.
  [arXiv:1011.5593 [gr-qc]].

%\cite{Barcelo:2006uw}
\bibitem{Barcelo:2006uw}
  C.~Barcelo, S.~Liberati, S.~Sonego, M.~Visser,
  %``Hawking-like radiation does not require a trapped region,''
  Phys.\ Rev.\ Lett.\  {\bf 97 } (2006)  171301.
  [gr-qc/0607008].

%\cite{Barbado:2011ai}
\bibitem{Barbado:2011ai}
  L.~C.~Barbado, C.~Barcelo, L.~J.~Garay and G.~Jannes,
  %``The Trans-Planckian problem as a guiding principle,''
  JHEP {\bf 1111} (2011) 112
  [arXiv:1109.3593 [gr-qc]].
  %%CITATION = ARXIV:1109.3593;%%

\end{thebibliography}
\end{document}